\documentclass[twoside]{dis07}
\usepackage[latin1]{inputenc}
\usepackage[dvips]{graphicx,epsfig,color}
\usepackage{wrapfig,rotating}
\usepackage{amssymb,amsmath,array}

\begin{document}
\title{The Role of Gauge Invariance in Single-Spin Asymmetries\protect\footnote{Talk presented at the 15th International Workshop on Deep-Inelastic Scattering and Related Subjects (DIS 2007), 
April 16-20, 2007, Munich, Germany.}}

\author{C.J.~Bomhof
and P.J.~Mulders
\vspace{.3cm}\\
Vrije Universiteit Amsterdam - Department of Physics and Astronomy \\
NL-1081 HV Amsterdam - the Netherlands}

\maketitle

\begin{abstract}
We argue that through the Wilson lines,
gauge invariance has as an effect that the hard functions in weighted spin-asymmetries in hadronic scattering processes are given by gluonic pole cross sections,
rather than the usual partonic cross sections.
\end{abstract}

\section{Introduction}

The interest in the transverse polarization properties of hadrons has been increasing considerably over the last decades after the observation of large single transverse-spin asymmetries $A_N$ in processes such as
$pp{\rightarrow}\Lambda^{0\uparrow}X$, $p^\uparrow p{\rightarrow}\pi X$ and 
$p^\uparrow\bar p{\rightarrow}\pi X$, see \emph{e.g.}~\cite{one}.
Typically, single-spin asymmetries (SSA) are defined as the difference of the scattering cross sections~with~op\-po\-site spin orientations divided by the spin averaged cross section.
As such, the spin asymmetries are pre-eminent observables for the spin dependent parton distribution functions,
such as the quark transversity function $h_1(x)$,
which cannot be probed at leading order in inclusive deep-inelastic scattering.

A mechanism to generate single-spin asymmetries through soft gluon interactions between the target remnants and the initial and final state partons was first proposed in the context of collinear factorization~\cite{three,nineteen}.
Hence, this collinear factorization formalism involves, 
apart from the usual twist-two collinear distribution functions,
also twist-three collinear quark-gluon matrix elements.
Since they contain a gauge field operator corresponding to a soft gluon, 
they are referred to as \emph{gluonic pole matrix elements}.
An important example is the Qiu-Sterman matrix element $T_F(x{,}x)$~\cite{three}.
Several other mechanisms for generating SSA's through the effects of the intrinsic transverse momenta of the partons have also been proposed.
For instance, in the Sivers effect the asymmetry arises in the initial state due to a correlation between the intrinsic transverse motion of an unpolarized quark and the transverse spin of its parent hadron~\cite{four}.
The effect is described by a transverse momentum dependent (TMD) function $f_{1T}^\perp(x{,}p_T^2)$.
However, based on time-reversal arguments it was expected to vanish~\cite{five}.
It will, therefore, be understood that it came as a surprise when Brodsky, Hwang and Schmidt in 2002 showed that leading twist spin asymmetries can anyway be generated in 
a spectator diquark model where the scattered quark undergoes an additional gluonic interaction with the target remnants~\cite{six}.
It was soon realized that the time-reversal arguments that were used to predict a vanishing Sivers effect were flawed as they neglected the presence of the Wilson lines.
The \emph{Wilson lines} or \emph{gauge links}
$\mathcal U\,{=}\,\mathcal P\exp[-ig{\int}dz{\cdot}A(z)]$ are path-ordered exponentials that are required to obtain gauge-invariant definitions of parton distribution and fragmentation functions.
Taking the Wilson lines properly into account it was found instead that time-reversal leads to the important conclusion that the Sivers functions in SIDIS and Drell-Yan scattering have opposite signs 
\begin{equation}\label{TMDsignflip}
f_{1T}^{\perp}(x{,}p_T^2)\;\big\rfloor_{\text{Drell-Yan}}
=-f_{1T}^{\perp}(x{,}p_T^2)\;\big\rfloor_{\text{SIDIS}}\ ,
\end{equation}
and that finite Sivers contributions are possible after all~\cite{seven}.
Single-spin effects due to the Sivers mechanism in SIDIS have now been observed by several collaborations~\cite{eight}.

The discussion above exemplifies the important role that the Wilson lines have come to play.
Today it is recognized that they are not just mere operators to
obtain gauge invariant definitions of parton distribution functions but also that they,
through the Sivers (and Boer-Mulders~\cite{nine}) effect,
can themselves be regarded as sources for single-spin asymmetries.
The Wilson lines are also crucial ingredients in the derivation of the relation $2Mf_{1T}^{\perp(1)}(x)\,{=}\,{-}gT_F(x{,}x)$ found in ref.~\cite{ten}, 
demonstrating that the first transverse moment of the Sivers function is a gluonic pole matrix element
(the same is true for the Boer-Mulders function).
The remarkable `universality' property of the TMD Sivers functions in Eq.~\eqref{TMDsignflip} 
is a prediction intrinsic to QCD which follows from the presence of the Wilson lines or, stated differently,
from the gluonic initial and final-state interactions.
Experimental verification of this prediction would be profound support for our understanding of the physics underlying the generation of spin asymmetries.
It could be tested by comparing the signs of the single transverse-spin asymmetries in SIDIS and Drell-Yan scattering.
As stated in the previous paragraph, 
SSA due to the Sivers mechanism in SIDIS have now been observed by several collaborations.
However, measurements of the single-spin asymmetry in Drell-Yan scattering are lacking behind.
The reason is that the lepton-antilepton pair is a relatively rare final state in hadron-hadron scattering compared to purely hadronic or hadron-photon final states.
Also in those processes the gluonic initial and final-state interactions 
(the Wilson lines) leave their fingerprints,
though the effects are more intricate since the hard functions are more complicated.
However, until precise measurements for the single-spin asymmetries in the `gold-plated' Drell-Yan process become available,
they may 
also be used to test the formalism describing single-spin asymmetries.
The role of gauge invariance in these more complicated scattering processes is the topic of the next section.

\section{Dijet and Photon-Jet Production in Hadron-Hadron Scattering}

In the basic hadronic processes, SIDIS, Drell-Yan scattering and $e^+e^-$-annihilation,
the hard functions are just simple electromagnetic vertices (at tree-level).
Depending on the particular process only initial or final-state gluonic interactions contribute and, correspondingly,
only future and past pointing Wilson lines occur.
However, when going to hadronic processes that involve hard functions with more colored external legs,
such as in hadronic dijet or photon-jet production,
there can be both initial and final state gluonic interactions.
As a result, the Wilson lines will also be more complicated than just the simple future and past pointing Wilson lines~\cite{thirteenA}.
In fact, the final-state interactions will give rise to future pointing Wilson lines at each of the outgoing partons 
(where the representations of the color-matrices will depend on the particular type of parton) 
as is the case in SIDIS.
Similarly, the initial-state interactions will lead to past pointing Wilson lines at the incoming partons as in Drell-Yan scattering.
The Wilson lines distributed over the different external partons of the hard function can subsequently be joined together by making a color-flow decomposition of the hard function,
such that they can be connected along the color-flow lines.
In particular this means that the full gauge link will have a different substructure for each of the color-flow diagrams.
More importantly, it implies that each of the Feynman diagrams that contribute to the hard function might have a different gauge link structure~\cite{thirteenA}.
For the TMD distribution functions this at first sight seems to complicate things considerably.
However, for the collinear distribution functions remarkable simplifications occur.
In fact, for the collinear $T$-even parton distribution functions all process dependence of the Wilson lines disappears.
For the $T$-odd distribution functions the effect of the jungle of Wilson lines is to relate their first transverse moments for different processes by simple color factors:
\begin{equation}\label{Rel}
f_{1T}^{\perp(1)}(x)\,\big\rfloor_{D(ab\rightarrow cd)}
=C_G^{[D(ab\rightarrow cd)]}\,
f_{1T}^{\perp(1)}(x)\,\big\rfloor_{\text{SIDIS}}\ ,
\end{equation}
and similarly for the transverse moments of the Boer-Mulders functions.
The color factors $C_G$ are determined by the gauge links and, hence, 
depend on the scattering process $ab{\rightarrow}cd$.
Moreover, they depend on the particular Feynman diagram $D$ that contributes to this process~\cite{thirteenA,thirteenB}.
For that reason they are in a natural way associated to the hard functions in which they can, then, be absorbed.
Consequently, the resulting hard functions will have the generic structure
$d\hat\sigma_{[a]b\rightarrow cd}\,
{=}\sum\nolimits_DC_G^{[D]}d\hat\sigma_{ab\rightarrow cd}^{[D]}$,
which will be referred to as \emph{gluonic pole cross sections} and which should be contrasted to the usual partonic cross sections
$d\hat\sigma_{ab\rightarrow cd}\,
{=}\sum\nolimits_Dd\hat\sigma_{ab\rightarrow cd}^{[D]}$.
The summations run over all Feynman diagrams $D$ that contribute to the scattering process $ab{\rightarrow}cd$ and the 
$d\hat\sigma^{[D]}$ are their pQCD expressions.
The gluonic pole cross section has a bracketed subscript indicating which parton is associated with the gluonic pole matrix element.
They appear as the hard functions whenever gluonic pole matrix elements contribute,
such as the first moments of the Sivers or Boer-Mulders functions.
This is typically the case in weighted azimuthal spin asymmetries.
Hence, the effect of the gauge links 
(or, equivalently, of the initial and final state interactions)
for weighted SSA's is that the hard functions of these observables will be given by the (manifestly gauge invariant) gluonic pole cross sections~\cite{thirteenB}, 
rather than the usual partonic cross sections as might be expected in a `generalized parton model approach'.
Several other recent theoretical studies of single-spin asymmetries seem to point in a similar direction~\cite{eleven,elevenb,twelve}, though more research is required to convey the exact connection between the different formalisms.

The effects of the gluonic initial and final-state interactions for the fully TMD treatment of these processes is not so clear-cut.
In ref.~\cite{twelve} a TMD factorization formula was proposed for the quark-Sivers contribution to the SSA in dijet production in proton-proton scattering.
This formula involves the gluonic pole cross sections found in refs.~\cite{thirteenB} as hard functions, 
but folded with the TMD distribution functions as measured in SIDIS
(\emph{i.e.}\ with a future pointing Wilson line in their definitions).
On the other hand, in refs.~\cite{thirteenA,thirteenB} it was observed that complicated Wilson line structures occur in the TMD distribution 
(and fragmentation) 
functions in such processes.
The work of~\cite{thirteenA,thirteenB} and~\cite{twelve} could be related by extending the relation in Eq.~\eqref{Rel} between the first moments of the Sivers functions to the full TMD functions,
in analogy to the relation between the Sivers functions in SIDIS and Drell-Yan scattering as given in Eq.~\eqref{TMDsignflip}.
However, the jungle of Wilson lines found in refs.~\cite{thirteenA,thirteenB} in concurrence with a model calculation
led the authors of ref.~\cite{coll} to conclude that a TMD factorization formula for spin asymmetries in processes such as dijet production in proton-proton scattering is not possible with universal distribution functions.
It is also asserted that a proof of TMD factorization for such processes will be essentially different from the existing proofs for SIDIS and Drell-Yan scattering and that it will possibly involve `effective' distribution functions~\cite{elevenb}.
Future research will have to clarify what the relation is between the results of refs.~\cite{thirteenA,thirteenB,twelve,coll}.
In particular, soft factors might play an important role in such a comparative study.

Using the azimuthal imbalance of the outgoing jet pair 
($p^\uparrow p{\rightarrow}jjX$) or photon-jet pair
($p^\uparrow p{\rightarrow}\gamma jX$) weighted single-spin asymmetries have been constructed that involve the gluonic pole cross sections at leading twist~\cite{thirteenB,fifteen,sixteen}.
Predictions for the quark-Sivers contribution to these spin-asymmetries in dijet production have been presented~\cite{url,fifteen},
making use of recently obtained parametrizations~\cite{seventeen} for the quark-Sivers first-moments (Qiu-Sterman matrix elements) obtained by fitting to $A_N$ data taken at E704, BRAHMS and STAR,
and for the quark-Sivers half-moments obtained by fitting to SSA's measured at HERMES.
First experimental measurements for this process have been performed at RHIC in 2006~\cite{fourteen}.
The results seem to indicate that the full QCD treatment,
which makes use of gluonic pole cross sections to account for the competing effects of the gluonic initial and final-state interactions, 
is consistent with the data,
while the `generalized parton model approach',
which only uses partonic cross sections, overestimates the measurements~\cite{fourteen}.
However, the experimental results are at the same time consistent with vanishing SSA's in dijet production and more measurements to increase statistics and disentangle flavor dependence in concurrence with more elaborate theoretical studies would therefore be welcome.

Predictions for the single-spin asymmetry in the similar photon-jet production 
process have been made in kinematical regions accessible at RHIC.
For these predictions use was made of parametrizations~\cite{eighteen} for the quark-Sivers function obtained by fitting to HERMES and COMPASS data.
The main result is that the effect of the gluonic initial and final-state interactions is to flip the sign of the spin-asymmetry in the full QCD treatment compared to the generalized parton model result~\cite{url,sixteen}.
This can be understood intuitively by observing that the dominant partonic channel is $qg{\rightarrow}\gamma q$ scattering,
where in the large-$N$ limit the color flows from the initial quark and through the gluon back into the initial state.
There the initial-state interactions lead to the sign flip of the spin asymmetry,
as is also the case in Drell-Yan scattering.
However, in the $qg{\rightarrow}\gamma q$ process the initial-state interactions are gluon self-couplings and, hence,
the experimental measurement of the predicted sign flip in this case would test the very non-abelian nature of QCD.
It is also shown that the Boer-Mulders and gluon-Sivers effects are too small to affect this conclusion.
Therefore, by measuring the sign of the asymmetry one has a test of the QCD formalism to describe the generation of single-spin asymmetries that is equally legitimate as the measurement of the SSA in Drell-Yan scattering.\\[0mm]

Part of this work was supported by the foundation for Fundamental Research of Matter (FOM) and the National Organization for Scientific Research (NWO).

\begin{footnotesize}

\end{footnotesize}

\end{document}